\documentclass[twocolumn,prl,showpacs,amsfonts,amsmath,assymb,eufrak]{revtex4-1}
\usepackage{epsfig}
\usepackage{color}
\usepackage{bm}
\usepackage{braket}
\usepackage{graphicx}
\usepackage{amsthm}

\usepackage{calc}

\newcommand{\eq}[1]{\begin{equation} #1 \end{equation}}
\newcommand{\eqa}[2]{\begin{equation} #1 \label{#2} \end{equation}}
\newcommand{\balign}[1]{\begin{align} #1 \end{align}}

\newcommand{\mx}[1]{\begin{pmatrix}#1 \end{pmatrix}}
\newcommand{\figin}[4]
{\begin{figure}[tb]
\centering
\includegraphics[width= #1]{#2.pdf}
\caption{#3}
\label{f:#4}
\end{figure}}

\newcommand{\todayd}{\the\year/\the\month/\the\day}

\newcommand{\bib}{\bibitem}

\newcommand{\lmd}{\lambda}

\newcommand{\lb}{\label}
\newcommand{\nt}{\notag}
\newcommand{\Tr}{\mathrm{Tr}}

\newcommand{\bel}{\begin{easylist}}
\newcommand{\eel}{\end{easylist}}
\newcommand{\be}[1]{\begin{enumerate} #1 \end{enumerate}}

\newcommand{\eref}[1]{Eq.~\eqref{#1}}

\newcommand{\fref}[1]{Fig.~\ref{f:#1}}

\def \({\left(}
\def \){\right)}
\def \[{\left[}
\def \]{\right]}

\newcommand{\abs}[1]{\left|#1\right|}

\newcommand{\sumtwo}[2]%
{\mathop{\sum_{#1}}_{#2}}
\newcommand{\sumthree}[3]%
{\mathop{\mathop{\sum_{#1}}_{#2}}_{#3}}
\newcommand{\sumfour}[4]%
{\mathop{\mathop{\mathop{\sum_{#1}}_{#2}}_{#3}}_{#4}} 
\newcommand{\prodtwo}[2]%
{\mathop{\prod_{#1}}_{#2}}
\newcommand{\mintwo}[2]%
{\mathop{\min_{#1}}_{#2}}
\newcommand{\maxtwo}[2]%
{\mathop{\max_{#1}}_{#2}}
\newcommand{\maxthree}[3]%
{\mathop{\mathop{\max_{#1}}_{#2}}_{#3}}
\newcommand{\limtwo}[2]%
{\mathop{\lim_{#1}}_{#2}}
\newcommand{\suptwo}[2]%
{\mathop{\sup_{#1}}_{#2}}
\newcommand{\supthree}[3]%
{\mathop{\mathop{\sup_{#1}}_{#2}}_{#3}}
\newcommand{\supfour}[4]%
{\mathop{\mathop{\mathop{\sup_{#1}}_{#2}}_{#3}}_{#4}} 
\newcommand{\inftwo}[2]%
{\mathop{\inf_{#1}}_{#2}}
\newcommand{\infthree}[3]%
{\mathop{\mathop{\inf_{#1}}_{#2}}_{#3}}
\newcommand{\inffour}[4]%
{\mathop{\mathop{\mathop{\inf_{#1}}_{#2}}_{#3}}_{#4}} 
\newcommand\calN{{\cal N}}

\newcommand{\bbR}{\mathbb{R}}

\newcommand{\ep}{\varepsilon}

\newcommand{\para}[1]{{\em #1}\/.---}
\def\rnum#1{\resizebox{0.5em}{\height}{\expandafter{\romannumeral #1}}}
\def\Rnum#1{\resizebox{0.5em}{\height}{\uppercase\expandafter{\romannumeral #1}}}

\newcommand{\rhoG}{\rho_{\rm Gibbs}}
\newcommand{\cG}{c_{\rm Gibbs}}
\newcommand{\sgmG}{\sigma_{\rm Gibbs}}
\newcommand{\omgG}{\omega_{\rm Gibbs}}
\newcommand{\OmgG}{\Omega_{\rm Gibbs}}

\newcommand{\SH}{S_{\rm H}}

\newcommand{\tlsgm}{\tilde{\sigma}}
\newcommand{\pwG}{p_{\rm w}^{\rm G}}

\makeatletter
\renewcommand{\@cite}[1]{\textsuperscript{#1)}}
\makeatother

\begin{document}

\preprint{APS/123-QED}

\newcommand{\titlehere}{Quantum thermodynamics of correlated-catalytic state conversion at small-scale}

\newcommand{\titlename}{\titlehere}

\preprint{APS/123-QED}

\title{\titlename}% Force line breaks with \\

\author{Naoto Shiraishi$^{1}$ and Takahiro Sagawa$^{2}$ }
\affiliation{$^{1}$Department of Physics, Gakushuin University, 1-5-1 Mejiro, Toshima-ku, Tokyo 171-8588, Japan} 

\affiliation{$^{2}$Department of Applied Physics and Quantum-Phase Electronics Center (QPEC),  The University of Tokyo, 7-3-1 Hongo, Byunkyo-ku, Tokyo 113-8656, Japan}%

\date{\today}% It is always \today, today,

\begin{abstract}
The class of possible thermodynamic conversions can be extended by introducing an auxiliary system called catalyst, which assists state conversion while remaining its own state unchanged.
We reveal a complete characterization of catalytic state conversion in quantum and single-shot thermodynamics by allowing an infinitesimal correlation between the system and the catalyst.
Specifically, we prove that a single thermodynamic potential, which provides the necessary and sufficient condition for the correlated-catalytic state conversion, is given by the standard nonequilibrium free energy defined with the Kullback-Leibler divergence.
This resolves the conjecture raised by Wilming, Gallego, and Eisert [Entropy 19, 241 (2017)]  and by Lostaglio and M\"{u}ller [Phys. Rev. Lett. 123, 020403 (2019)] in positive.
Moreover, we show that, with the aid of the work storage, any quantum state can be converted into another by paying the work cost equal to the nonequilibrium free energy difference.
Our result would serve as a step towards establishing resource theories of catalytic state conversion in the fully quantum regime.
\end{abstract}

\maketitle

%%%%%%%%%%

\para{Introduction}
The extension of thermodynamics to small-scale quantum systems has attracted attention in various research fields.
Variety of the second laws employing the R\'{e}nyi entropies and divergences~\cite{AN, Tur, Kli, Bra15} or majorization~\cite{MOAbook, Gour-review, Los-review, Sag-review} naturally arise in the small-scale, which is contrastive to conventional thermodynamics where only a single thermodynamic potential such as the equilibrium free energy characterizes state convertibility~\cite{LY}.
Recent studies pushing toward this direction are developed in terms of resource theories~\cite{CG19, Gour-review, Los-review}.
The resource theory of athermality~\cite{Jan00, Bra13, Bra15, HO} paves the way for establishing the information-theoretic foundation of thermodynamics.

In resource theories, an auxiliary system called {\it catalyst} plays a key role~\cite{JP99}, assisting the state conversion while the catalyst itself does not change.
To formulate the catalytic state conversion, we suppose the composite system of the system and the catalyst, and consider a state conversion $\rho\otimes c\to \sigma\otimes c$, where $\rho$, $\sigma$ are states of the system and $c$ is a state of the catalyst.
On one hand, if we require the exact return of the catalyst, an infinite family of R\'{e}nyi entropies or divergences characterizes possible catalytic state conversion~\cite{AN, Tur, Kli, Bra15}.
On the other hand, if we allow a small finite error in the final state of the catalyst, any state conversion is possible, which is called {\it embezzling}~\cite{Bra15, DH, Ng15}.
Here our focus lies in their intermediate regime, where another nontrivial characterization of state convertibility emerges.

Specifically, we consider the situation that the catalyst returns to its initial state exactly but with a negligibly small correlation between the system and the catalyst.
As observed in Refs~\cite{MP, LMP}, stochastic independence (absence of correlations) is a resource of thermodynamic state conversions.
Along with this idea, Wilming, Gallego, and Eisert~\cite{WGE} conjectured that the nonequilibrium free energy defined by the quantum Kullback-Leibler (KL) divergence gives the unique criterion of correlated-catalytic state conversion via a Gibbs-preserving map with a negligibly small correlation.
In the classical case, this conjecture has been solved in positive by M\"{u}ller~\cite{Mul} and generalized by Rethinasamy and Wilde~\cite{RW}.
However, these results cannot apply to the quantum case, because unlike the classical case known criteria for quantum relative majorization are highly complicated~\cite{BG17, Bus, Gou}.
Therefore, the original conjecture raised in Ref~\cite{WGE} (also raised in Ref.~\cite{LM19} in a rigorous manner) for the quantum cases has still been left as a highly-nontrivial open problem.

In this Letter, we solve this problem for the quantum case~\cite{WGE, LM19} in the affirmative:
We prove that the KL divergence indeed characterizes quantum correlated-catalytic state conversion in a necessary and sufficient manner.
That is, the correlated-catalytic state conversion between two given quantum states by a Gibbs-preserving map is possible {\it if and only if} the nonequilibrium free energy defined by the KL divergence does not increase.
We further prove that even if the final free energy is larger than the initial one, we can still convert the initial state to the final one by adding a two-level work storage and paying the work cost equal to or greater than the free energy difference.
Our result implies that the conventional form of the second law given by the KL divergence is restored even in the quantum regime, if the catalyst is allowed to correlate with the system.
%We thus regard our result as a conceptual foundation that bridges conventional thermodynamics and the resource theory.
%Furthermore, our result would have implications not only to quantum thermodynamics but also to various quantum resource theories~\cite{CG19}, as our proof techniques apply to quantum state conversion that is not necessarily thermodynamic.

\para{Setup and the main claim}
Consider a finite-dimensional quantum system with Hamiltonian $H$.
We investigate state conversion through a particular class of the completely-positive and trace-preserving (CPTP) maps, called {\it Gibbs-preserving maps} $\Lambda$, which keep the Gibbs state invariant: $\Lambda (\rho_{\rm Gibbs})=\rho_{\rm Gibbs}$. 
Here, $\rhoG:=e^{-\beta H}/Z$ is the Gibbs state with the inverse temperature $\beta$ of the environment.
We set the Boltzmann constant to unity.
In terms of the resource theory of athermality, the Gibbs state is a free state (with zero athermality), and Gibbs-preserving maps do not generate any non-free state (with nonzero athermality) from a free state.

We employ an external system called catalyst denoted by $C$, which assists state conversion of the system $S$ while the state of $C$ itself does not change (see also \fref{schematic}).
As in Refs.~\cite{Bra13, Bra15, MNW, MP, LMP, Mul, RW, Fai, Sag}, we allow a negligibly small error on the final state of $S$, while the marginal state of $C$ exactly goes back to the initial state.
The most crucial assumption is to allow a negligibly small correlation between $S$ and $C$ in the final state.
This assumption is motivated by the fact that negligibly small correlations are always allowed between the system and the environment in conventional thermodynamics.
In terms of resource theories, a catalyst for a system can be reused as a catalyst for other systems even when a correlation with the first system remains.

\figin{8.5cm}{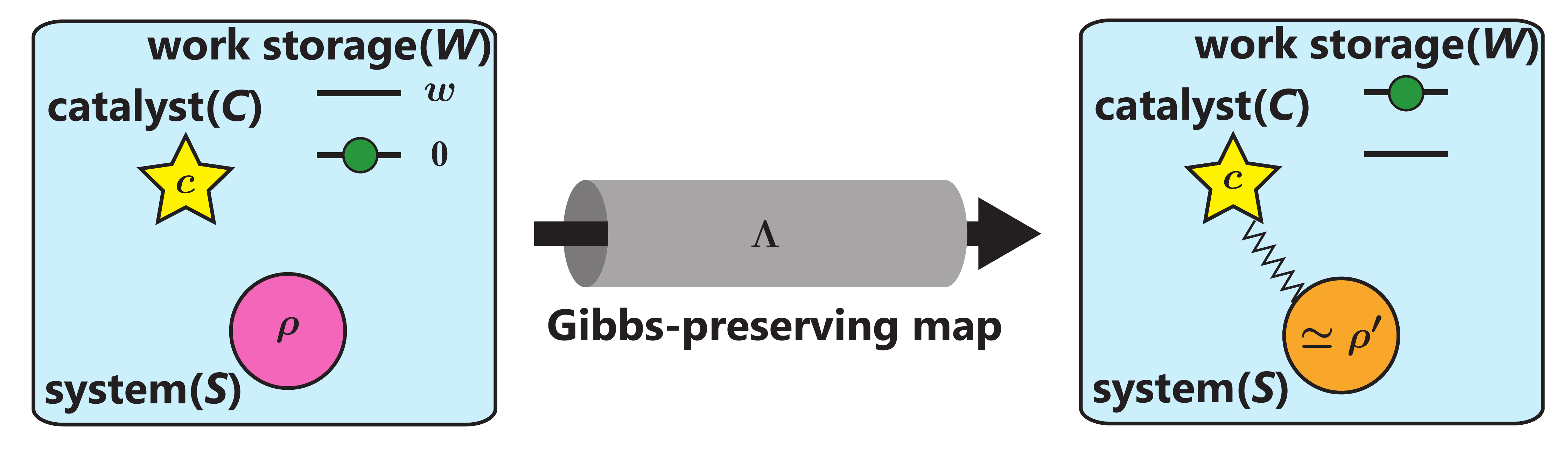}{
Schematic of our setup.
We convert the system $S$ from $\rho$ to $\rho'$ with the aid of the catalyst $C$ and the work storage $W$.
The catalyst $C$ returns to its original state while it can correlate with the system.
The work storage $W$ changes its state with energy difference $w\leq F(\rho)-F(\rho')$ with probability arbitrarily close to unity.
}{schematic}

We define the nonquilibrium free energy as $F(\rho):=S_1(\rho||\rhoG)$, where $S_1(\rho||\rhoG):=\Tr[\rho \ln \rho]-\Tr[\rho \ln \rhoG]$ is the KL divergence~\cite{NCbook}.
We now state our first main theorem:

\smallskip

{\it Theorem 1}--
Consider two quantum states of $S$; $\rho$ and $\rho'$.
Then, $F(\rho)\geq F(\rho')$ is satisfied if and only if there exist a catalyst $C$ and its state $c$, and a Gibbs-preserving map $\Lambda$ satisfying $\Lambda(\rho \otimes c)=\tau$ such that (i) $\Tr_S[\tau]=c$, (ii) $\Tr_C[\tau]$ is arbitrarily close to $\rho'$, (iii) the correlation between $S$ and $C$ in the final state is arbitrarily small.

\smallskip

The fully rigorous statement of the above theorem and its proof are presented in Supplemental Material~\cite{Supple}.
Here, we only remark that the closeness between states is quantified by the trace distance $d_1(\rho', \rho''):=\frac12 \Tr[\abs{\rho'-\rho''}]$ and the amount of the correlation is quantified by the mutual information $I_{\rm SC}[\tau]:=S_1(\tau||\rho''\otimes c)$, where $\rho''=\Tr_C[\tau]$ is the reduced state of $\tau$ on $S$.
An arbitrarily small error can be achieved by choosing an appropriate catalyst, which might be very large.
This theorem manifests that the free energy $F(\rho)$ serves as the single monotone of quantum thermodynamics at the small-scale if we allow a negligibly small correlation between the system and the catalyst.

In the case of $F(\rho)<F(\rho')$, Theorem 1 implies that we cannot convert $\rho$ to $\rho'$ through any Gibbs-preserving map.
However, even in this case, we can convert $\rho$ to $\rho'$ with the aid of the work storage $W$ (see \fref{schematic}).
The work storage is a two-level system which compensates for the energy change in $S$ by investing the work cost.
The initial state of $W$ is an energy eigenstate $\ket{a}$ with energy $E_a$, and the final state is arbitrarily close to another energy eigenstate $\ket{b}$ with energy $E_b$.
Thus, the work value is almost deterministic, which is an approximate version of the single-shot scenario~\cite{Mul, Sag-review, WNW}.

By applying Theorem 1 to the composite system $SW$, we find that $\rho\otimes \ket{a}\bra{a}$ can be converted to a state close to $\rho'\otimes \ket{b}\bra{b}$ with a catalyst if we allow a correlation between $SW$ and $C$.
Further to that, we can prove a much stronger statement: the desired state conversion is possible even when there is no correlation between $W$ and the remaining part $SC$:

\smallskip

{\it Theorem 2}--
Consider two quantum states $\rho$ and $\rho'$ of the system $S$ with $F(\rho)-F(\rho')<0$.
Then, $F(\rho)-F(\rho')\geq w$ is satisfied if and only if there exist a catalyst $C$ and its state $c$, a work storage $W$ with $E_b-E_a\geq w$, and a Gibbs-preserving map $\Lambda$ satisfying $\Lambda(\rho \otimes c \otimes \ket{a}\bra{a})=\tau \otimes \omega$ with $\tau$ and $\omega$ being states of $SC$ and $W$, such that (i) $\Tr_S[\tau]=c$, (ii) $\Tr_C[\tau]$ is arbitrarily close to $\rho'$, (iii) $\omega$ is arbitrarily close to $\ket{b}\bra{b}$, (iv) the correlation between $S$ and $C$ in $\tau$ is arbitrarily small.

\smallskip

This theorem reveals the minimum work cost when $C$ correlates only with $S$ as depicted in \fref{schematic}, and represents {\it the principle of maximum work}~\cite{Fermibook,Callenbook,  Landaubook}.
The foregoing two theorems together provide the second law of quantum thermodynamics in the small-scale, yet in the apparently same form as conventional macroscopic thermodynamics.

We note that Theorem 2 only applies the case of the work investment ($w<0$), and does not cover the case of the work extraction ($w>0$).
We will, however, discuss a sufficient condition for the case of work extraction in Supplemental Material (Lemma 3)~\cite{Supple}.

\para{Outline of the proof}
The {\it if} part is easy to prove by applying superadditivity of the KL divergence.
Therefore, we here summarize the outline of the proof of the {\it only if} part.
The detailed idea is demonstrated along with a simple example soon later, and the full proofs are presented in Supplemental Material~\cite{Supple}.
We mainly treat Theorem 1 and briefly comment on Theorem 2.

Our proof consists of three steps.
%: deriving a sufficient condition for quantum state conversion by a Gibbs-preserving map (Step 1), applying the quantum Stein's lemma (Step 2), and the reduction from asymptotic state conversion to catalytic state conversion (Step 3).
In Step 1, we provide a sufficient condition to convert a quantum state $\sigma$ to another state $\sigma'$ via a Gibbs-preserving map by explicitly constructing the desired map.
We first perform a binary quantum measurement to determine whether the state is $\sigma$ or the Gibbs state $\sgmG$, and then prepare two quantum states depending on the measurement outcome.
In the case of Theorem 2 (with work storage), we consider not binary but ternary measurements, which require more careful treatment.
The derived sufficient condition employs two kinds of divergences; the quantum hypothesis testing divergence~\cite{WR12} and the quantum R\'{e}nyi divergence.
%, whose rigorous definitions are provided in the explanation of the toy example below.
This sufficient condition has been obtained by some literature explicitly~\cite{BST} and implicitly~\cite{FR, Fai, Sag, Sag-review, WW19}.

In Step 2, we apply the quantum Stein's lemma, which claims the convergence of the quantum hypothesis testing divergence rate to the KL divergence rate in the limit of infinitely many copies of given quantum systems~\cite{HP, ON}.
The reason why quantum hypothesis testing appears in quantum thermodynamics is that the $\ep$-smoothed R\'{e}nyi-$\infty$ divergence (introduced later) is bounded from both above and below by two quantum hypothesis testing divergences, and hence it also converges to the KL divergence~\cite{NH, Dat}.
%The quantum Stein's lemma plays the role to reduce various conditions with various divergences to a single condition with the Kullback-Leibler divergence.
In generic systems, the size of the catalyst diverges with vanishing error and correlation ($\ep, \delta\to 0$).
%A smaller error and correlation require a larger catalyst.
The necessary size of a catalyst can be evaluated by examining the aforementioned convergence speed~\cite{HP, ON, NH, Dat}.

In Step 3, we reduce the result on asymptotic state conversion (with multiple copies of states) of Step 1 and 2 to catalytic state conversion.
Although this type of reduction has been discussed in some literature~\cite{Dua, AN, Bra15}, we need some modification on the existing technique to keep the catalyst at the same state, because our asymptotic state conversion accompanies errors.
Combining these three steps, we arrive at the desired result.

\figin{8.5cm}{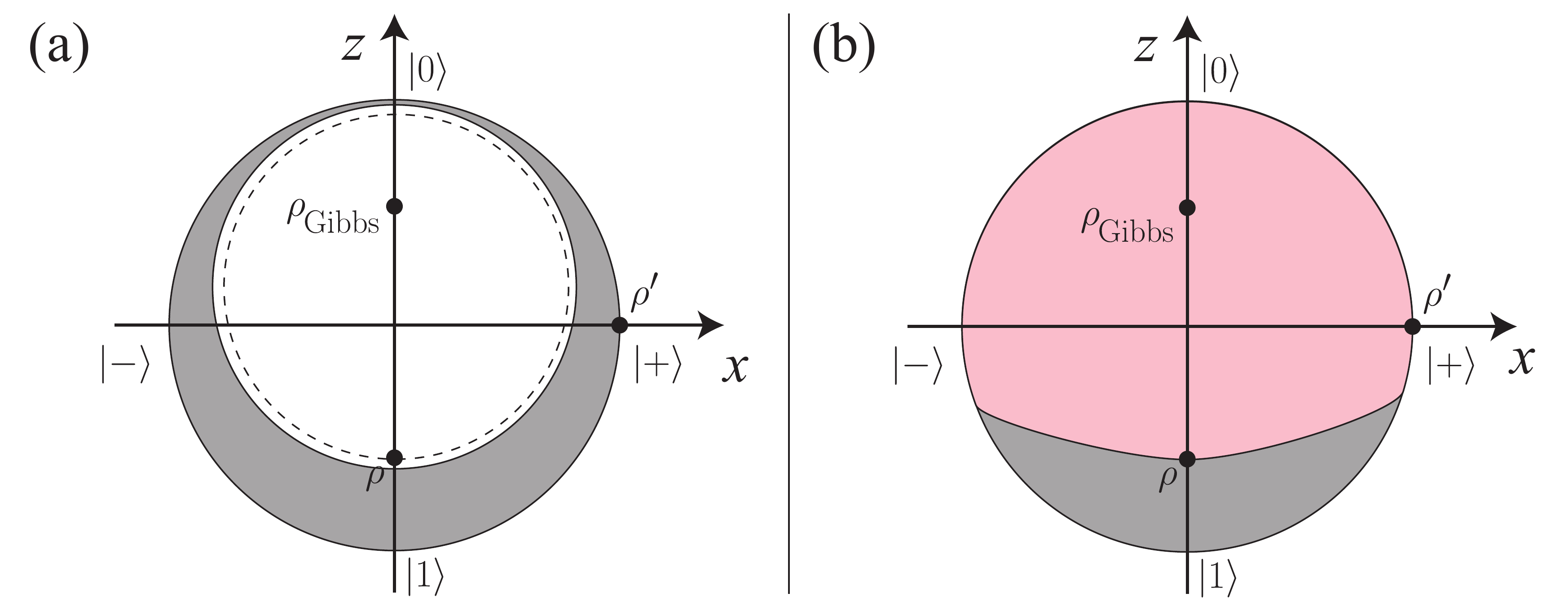}{
(a) Schematic of the criterion $S_{\infty}(\rho'||\rhoG)<S_\infty(\rho||\rhoG)$ represented by the dashed line in the $x$-$z$ plane of the Bloch sphere.
We draw the states inconvertible from $\rho$ within error $\ep$ in gray.
(b) Schematic of the criterion $S_1(\rho'||\rhoG)<S_1(\rho||\rhoG)$ in the $x$-$z$ plane of the Bloch sphere, where we draw the convertible and inconvertible states with a correlated catalyst from $\rho$ in red and gray, respectively.
In particular, there exists a Gibbs-preserving map with correlated catalyst converting $\rho\to \rho'$.
}{ex-Bloch}

\para{Toy example of Theorem 1}
We demonstrate the proof of the {\it only if} part of Theorem 1 in a toy example, a two-level system spanned by $\{ \ket{0}, \ket{1}\}$.
%The following argument also serves as a proof of Theorem 1 in this specific case.
We will construct two states that are not convertible from one to the other without catalyst, but are convertible with a correlated-catalyst.
For this purpose, we set $\rho=\frac{3}{200} \ket{0}\bra{0}+\frac{197}{200} \ket{1}\bra{1}$, $\rho'=\ket{+}\bra{+}$ with $\ket{+}:=\frac{1}{\sqrt{2}}(\ket{0}+\ket{1})$, $\beta=1$, $E_0=0$ and $E_1=\ln 3$.
The Gibbs state is given by $\rhoG=\frac34\ket{0}\bra{0}+\frac14\ket{1}\bra{1}$.
We set the upper bound of the error and the correlation strength as $\ep=0.01$ and $\delta=0.06$, respectively.
This state conversion is fully quantum because $\rho'$ is not diagonal in the energy eigenbasis.
We remark that any Gibbs-preserving map without catalyst cannot convert $\rho$ to $\rho'$.
To see this, we introduce the R\'{e}nyi-$\infty$ divergence $S_\infty (\sigma||\kappa):=\ln (\min[\lmd: \sigma\leq \lmd \kappa])$ and its $\ep$-smoothing $S^\ep_\infty (\sigma||\kappa):=\min_{d(\sigma', \sigma)\leq \ep}S_\infty (\sigma||\kappa)$ with $\ep>0$.
The R\'{e}nyi divergence satisfies the monotonicity under CPTP maps, and hence $S^\ep_{\infty}(\rho'||\rhoG)\leq S_\infty(\rho||\rhoG)$ is a necessary (but not sufficient) condition for state conversion without catalyst.
However, we can show $S^\ep_{\infty}(\rho'||\rhoG)>S_\infty(\rho||\rhoG)$ in the above parameter setting (see \fref{ex-Bloch}(a)).
We note that the above catalytic state conversion requires only 11 qubits, which is accessible by recent or near-term experimental techniques of e.g., superconducting qubits~\cite{IBM, Google}.

We treat Step 1 and Step 2 in parallel.
Consider a composite system of 8 copies of the two-level system: $\{ \ket{0}, \ket{1}\} ^{\otimes 8}$.
We construct a CPTP map which converts $\rho^{\otimes 8}$ to a state $\Xi$ satisfying $d_1(\Xi, {\rho'}^{\otimes 8})<\ep$ while keeping ${\rhoG}^{\otimes 8}$ unchanged.
We introduce the projection operator $Q$ onto the subspace of $\{\ket{0},\ket{1}\}^{\otimes 8}$ spanned by a subset of the computational basis that contains at most one $\ket{0}$.
We perform the binary measurement with $\{ Q, 1-Q\}$ in order to distinguish $\rho^{\otimes 8}$ and $\rhoG^{\otimes 8}$.
By this measurement, $\rho^{\otimes 8}$ outputs $Q$ with probability $0.994\cdots >1-\ep$, and $\rhoG^{\otimes 8}$ outputs $1-Q$ with probability $1-\frac{25}{4^8}$.
Their differences from 1 (i.e., $0.005\cdots$ and $\frac{25}{4^8}$) correspond to the error of the first and the second kinds~\cite{Supple}, respectively.
We then prepare quantum states depending on the measurement outcome.
If the outcome is $Q$, we prepare the state $\ket{+}\bra{+}$, and if the outcome is $1-Q$, we prepare the state $\zeta$ expressed as
\eq{
\zeta =\frac{1}{1-\frac{25}{4^8}}\( \rhoG^{\otimes 8} -\frac{25}{4^8}\ket{+}\bra{+}^{\otimes 8}\) ,
}
which is positive-semidefinite because $\rhoG^{\otimes 8}-\lmd \ket{+}\bra{+}^{\otimes 8}\geq 0$ for $\lmd\leq \( \frac{3}{8}\) ^8$~\cite{Supple} and $\frac{25}{4^8}<\( \frac{3}{8}\) ^8$.
This measurement-and-preparation procedure indeed converts $\rhoG$ to $\rhoG$ by construction and converts $\rho$ to $\Xi:=(0.994\cdots)\ket{+}\bra{+}^{\otimes 8}+(0.005\cdots)\zeta$, which satisfies $d_1(\Xi, \ket{+}\bra{+}^{\otimes 8})<\ep$.
We denote this CPTP map by $\Lambda$.

We next move to Step 3.
We identify the system $S$ to $S_1$ and the catalyst $C$ to $S_2\otimes \cdots \otimes S_8\otimes A$, where $A$ is an auxiliary system spanned by a basis $\{ \ket{1}, \ket{2},\ldots , \ket{8}\}$.
The Hamiltonian of $A$ is set to be trivial (i.e., all the states in $A$ take the same energy).
Using $\Xi$ on $S_1\otimes\cdots\otimes S_8$ introduced above, we define $\Xi_i$ ($i=1,\ldots,8$) as the reduced state of $\Xi$ on $S_1\otimes\cdots\otimes S_i$.
We set $\Xi_0:=1\in\bbR$ (i.e., the trivial state) for convenience.
Using these states, we set the state of $C$ as 
\eq{
c:=\frac18 \sum_{k=1}^8 \rho^{\otimes k-1}\otimes \Xi_{8-k}\otimes \ket{k}\bra{k},
}
where $\rho^{\otimes k-1}$ is the state of $S_2\otimes\cdots\otimes S_k$, and $\Xi_{8-k}$ is now the state of $S_{k+1}\otimes\cdots\otimes S_8$.
The initial state of the composite system is $\rho\otimes c =\frac18 \sum_{k=1}^8 \rho^{\otimes k}\otimes \Xi_{8-k}\otimes \ket{k}\bra{k}$ (see \fref{swap}(a)).
We now construct the desired CPTP map as follows:
If the auxiliary system $A$ is $\ket{8}\bra{8}$, then we apply $\Lambda$ to $S_1\otimes\cdots\otimes S_8$, and leave it unchanged otherwise.
%Otherwise, we change the state of $A$ from $\ket{l}\bra{l}$ to $\ket{l+1}\bra{l+1}$.
Then, we shift the auxiliary system $A$ as $\ket{8}\to \ket{1}$ and $\ket{n}\to \ket{n+1}$.
Through this process, $\frac18 \sum_{k=1}^8 \rho^{\otimes k}\otimes \Xi_{8-k}\otimes \ket{k}\bra{k}$ is converted into $\tau'=\frac18 \sum_{k=1}^8 \rho^{\otimes k-1}\otimes \Xi_{9-k}\otimes \ket{k}\bra{k}$ (see \fref{swap}.(b)).
Remarkably, the partial trace of $\tau'$ with respect to $S_8$ recovers the initial state of the catalyst $c$.
In addition, by defining $\xi_l$ as the reduced state of $\Xi$ on $S_l$, the reduced state of $\tau'$ on $S_8$ is expressed as $\frac 18 \sum_{l=1}^8 \xi_l$, which is $\ep$-close to the desired state $\rho'=\ket{+}\bra{+}$ because $d_1(\Xi, \ket{+}\bra{+}^{\otimes 8})<\ep$.
Moreover, since $\Lambda$ is a Gibbs-preserving map, the constructed CPTP map is also Gibbs-preserving.
Thus, swapping the two-level systems as $S_n\to S_{n+1}$ and $S_8\to S_1$ after the above CPTP map, we arrive at the desired Gibbs-preserving map: 
$\rho\otimes c$ is converted into $\tau$ with $\Tr_S[\tau]=c$, $d_1(\Tr_C[\tau], \rho')<\ep$, and $I_{\rm SC}(\tau)<\delta$.
Here, since $S$ is a two-level system, $d_1(\tau, \rho'\otimes c)<\ep$ implies $I_{\rm SC}(\tau)<-\ep \ln \ep-(1-\ep)\ln (1-\ep)=0.056\cdots <0.06$.

\figin{8.8cm}{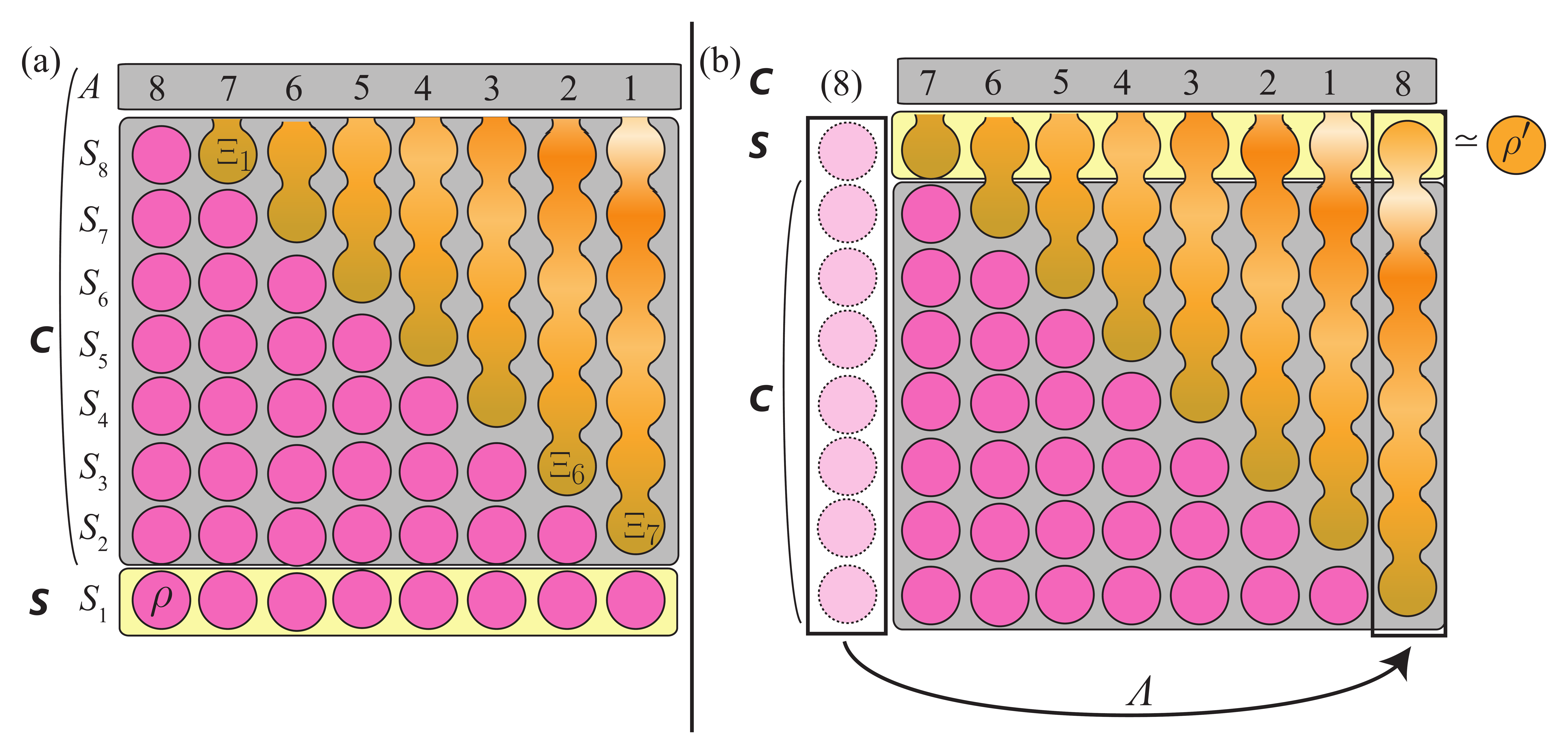}{
Schematic of Step 3 of the proof.
(a) The initial state of the composite $SC$.
The vertical direction represents different systems $S_1,\ldots , S_8$, and the horizontal direction means their classical mixture.
%We here draw only 6 two-level systems $S_1,\ldots ,S_6$, not $S_1, \ldots , S_8$ for brevity.
(b) Schematic of how the CPTP map $\Lambda$ gives the desired catalytic state conversion.
}{swap}

\para{Discussion}
The obtained results solve in positive the conjecture raised in Refs.~\cite{WGE, LM19}.
Note that M\"{u}ller~\cite{Mul} proved this conjecture for classical systems by showing an elaborate way to explicitly construct a catalyst, which is completely different from our approach.
Thus, our proof restricted to the classical regime serves as an alternative proof of M\"{u}ller's.

In this work, we have considered Gibbs-preserving maps as thermodynamic processes instead of thermal operations, while thermal operations are often regarded as proper operations in terms of the resource theory of thermodynamics.
These two classes of operations are equivalent in the classical regime~\cite{HO, Shi20}, while some Gibbs-preserving maps cannot be implemented by thermal operations in the quantum regime~\cite{FOR}.
The original conjecture of Ref.~\cite{WGE} is about Gibbs-preserving maps, and a stronger conjecture with quantum thermal operations was raised in Ref.~\cite{Mul}.
However, Refs.~\cite{LM19, MS19} solved the latter stronger conjecture in negative by proving that coherence cannot be broadcast.
In other words, no thermal operation converts an incoherent initial state to any coherent state even with the aid of correlated-catalyst.
This implies that one should consider a broader class of operations than thermal operations in order to enable characterization of state convertibility by the KL divergence.
In the present work, we focus on Gibbs-preserving maps that can still give a positive answer to the original conjectures~\cite{WGE, LM19}.
However, we expect that there may be a pathway to reduce Gibbs-preserving maps to thermal operations in the context of correlated-catalyst.
For example, as considered in Ref.~\cite{Fai}, the assistance of a small amount of coherence carried by an auxiliary system may enable such reduction in the asymptotic limit (i.e., the large catalyst limit in our setup).
Investigation of such a direction is an important future problem.

We also note that M\"{u}ller~\cite{Mul} performs trivialization of the catalyst Hamiltonian, but we did not.
Here, we say a catalyst trivialized when the Hamiltonian of the catalyst is trivial.
The hardness of trivialization in the fully quantum regime comes from the fact that merging and splitting states are irreversible due to decoherence in quantum systems.
Owing to this difficulty, we did not trivialize the catalyst in the present work.

Besides the correlated classical cases~\cite{MP, LMP, Mul, SCR}, in some setups of quantum thermodynamics and resource theories, a single thermodynamic potential with the KL divergence also appears~\cite{Bra15, Bra13, MNW, Fai, Sag, SSP13, Abe14, SSP14, Kor16}.
However, those previous results are different from our result in some important aspects:
Some of them allow small changes in the catalyst~\cite{Bra15} or other external systems~\cite{SSP13, Abe14, SSP14, Kor16} (instead they consider more restricted classes of operations compared to Gibbs-preserving maps), and some others consider asymptotic (macroscopic) conversion~\cite{Bra13, MNW, Fai, Sag}.
It is yet interesting to see that the same thermodynamic potential appears in these various setups.

Meanwhile, we can further extend Theorem 1 to general CPTP maps by employing a similar proof idea. 
This is about the quantum counterpart of catalytic d-majorization (also called relative majorization)~\cite{Gour-review, Sag-review}:

\smallskip

{\it Theorem 3}--
Consider four quantum states $\rho$, $\rho'$, $\eta$, and $\eta'$ of the system $S$ satisfying ${\rm supp}(\rho)\subset {\rm supp}(\eta)$.
Then, $S(\rho||\eta)\geq S(\rho'||\eta')$ holds if and only if there exists a catalyst $C$ and its two states $c$, $d$ satisfying ${\rm supp}(c)\subset {\rm supp}(d)$, and a CPTP map which converts $\eta \otimes d$ to $\eta'\otimes d$ and $\rho\otimes c$ to $\tau$ satisfying (i) $\Tr_S[\tau]=c$, (ii) $\Tr_C[\tau]$ is arbitrarily close to $\rho'$, (iii) the correlation between $S$ and $C$ in $\tau$ is arbitrarily small.

\smallskip

We present the proof of this theorem in Supplemental Material~\cite{Supple}.
This theorem almost solves the conjecture of Rethinasamy and Wilde~\cite{RW}, who proved the classical case.
The only difference between our result and the conjecture is that we did not trivialize catalyst $c$.

Meanwhile, Brandao and Gour~\cite{BG} have established that various resource theories concerning asymptotic state conversion with a small error are characterized by the KL divergence.
Their result applies to the resource theories of entanglement, coherence, contextuality, and stabilizer computation.
By employing our technique (in particular, Step 3 of the proof), we see that the KL divergence also serves as a single monotone in these single-shot resource theories with a correlated catalyst.
More generally, our approach developed in the present paper sheds new light on single-shot resource theories with catalyst, as recently demonstrated in Ref.~\cite{Wil20}.

%From \para{Intro} to here: Word length 3494 (ver5-3)

%If SM remains, 112 words is overestimated.

\bigskip

\para{Acknowledgement}
Authors thank Mark M. Wilde, Markus M\"{u}ller, and Francesco Buscemi for their careful reading and helpful comments.
Authors also thank R. Rubboli, M. Tomamichel, and T. Nagasawa for pointing out our improper descriptions.
NS is supported by JSPS KAKENHI Grants-in-Aid for Early-Career Scientists Grant Number JP19K14615. 
TS is supported by JSPS KAKENHI Grant Numbers JP16H02211 and JP19H05796.
TS is also supported by institute of AI and Beyond of the University of Tokyo.

%present: Acknowledge has 57 words (ver7-2) / 55 words (ver7)

%From \para{Intro} to here: Word length 3910 by APS (ver7) including Acknowledgement

\clearpage

\pagestyle{empty}

%%%%%%%%%%%%%%%%%%%%%%%%%%%%%%%%%%%%%%%%%%%
%%%%%%%%%%%%%%%%%%%%%%%%%%%%%%%%%%%%%%
% To modify figure captions
\makeatletter
\long\def\@makecaption#1#2{{
\advance\leftskip1cm
\advance\rightskip1cm
\vskip\abovecaptionskip
\sbox\@tempboxa{#1: #2}%
\ifdim \wd\@tempboxa >\hsize
 #1: #2\par
\else
\global \@minipagefalse
\hb@xt@\hsize{\hfil\box\@tempboxa\hfil}%
\fi
\vskip\belowcaptionskip}}
\makeatother
%%%%%%%%%%%%%%%%%%%%%%%%%%%%%%%%%%%%%%
\newcommand{\vo}{\upsilon}
\newcommand{\midskip}{\vspace{3pt}}

\setcounter{equation}{0}
\def\theequation{A.\arabic{equation}}
%%%%%%%%%%%%%%%%%%%%%%%%%%%%%%%%%%%%%%%

\begin{widetext}

\begin{center}
{\large \bf Supplemental Material for  \protect \\ 
  ``\titlehere'' }\\
\vspace*{0.3cm}
Naoto Shiraishi$^{1}$ and Takahiro Sagawa$^{2}$ \\
\vspace*{0.1cm}

{$^{1}$Department of Physics, Gakushuin University, 1-5-1 Mejiro, Toshima-ku, Tokyo 171-8588, Japan} 

{$^{2}$Department of Applied Physics and Quantum-Phase Electronics Center (QPEC),  The University of Tokyo, 7-3-1 Hongo, Byunkyo-ku, Tokyo 113-8656, Japan}%
\end{center}

\setcounter{equation}{0}
\renewcommand{\theequation}{S.\arabic{equation}}

In this Supplemental Material, we provide rigorous proofs of the theorems presented in the main text.
This Supplemental Material has the reference numbers in common with the main text.

%\begin{thebibliography}{99}
%\end{thebibliography}

%%%%%%%%%%%%%%%%%%%%%%%%%%%%%%%%%%%%%%%
\bigskip\noindent
{\bf \large A. Proof of Theorem 1 and Theorem 3}
\midskip
%%%%%%%%%%%%%%%%%%%%%%%%%%%%%%%%%%%%%%%

We first describe the fully rigorous statement of Theorem 3.
We note that Theorem 3 includes Theorem 1.

\bigskip

{\it Theorem 3}--
Consider four quantum states $\rho$, $\rho'$, $\eta$, $\eta'$ of the system $S$ satisfying ${\rm supp}(\rho)\subset {\rm supp}(\eta)$.
The following two are equivalent.
\be{
\renewcommand{\theenumi}{\arabic{enumi})}
\item $S_1(\rho||\eta)\geq S_1(\rho'||\eta')$.
\item For any $\ep>0$ and $\delta>0$, there exist a catalyst system $C$, its states $c$ and $d$ satisfying ${\rm supp}(c)\subset {\rm supp}(d)$, and a CPTP map $\calN$ on the composite system $SC$ such that $\calN(\rho\otimes c)=\tau$ and $\calN(\eta\otimes d)=\eta'\otimes d$, where $\tau$ satisfies $\Tr_S[\tau]=c$, $d_1(\Tr_C[\tau], \rho')<\ep$, and $I_{SC}[\tau]<\delta$.
}

Here, $S_1(\rho||\eta):=\Tr[\rho \ln \rho-\rho\ln \eta]$ is the quantum Kullback-Leibler (KL) divergence, $d_1(\rho', \rho''):=\frac12 \Tr[\abs{\rho'-\rho''}]$ is the trace distance, and $I_{\rm SC}[\tau]:=S_1(\tau||\rho''\otimes c)$ is the mutual information, where $\rho''=\Tr_C[\tau]$ is a reduced state of $\tau$ on $S$.
In Theorem 3, the map $\calN$ exactly converts $\eta$ to $\eta'$ with a catalyst, and also converts $\rho$ to $\rho'$ with a catalyst within arbitrary accuracy.
By setting $\eta=\eta'=\rhoG$, and setting the catalyst Hamiltonian such that its Gibbs state is $d$, Theorem 3 reduces to Theorem 1 in the main text.
%This generalized version coincides with the setup of Rethinasamy and Wilde~\cite{RW}.

\begin{proof}
The proof of $2)\Rightarrow 1)$ is easy.
The additivity, superadditivity and the monotonicity of the KL divergence imply that
\eq{
S_1(\rho||\eta)+S_1(c||d)=S_1(\rho\otimes c||\eta\otimes d)\geq S_1(\tau||\eta'\otimes d)\geq S_1(\Tr_C[\tau]||\eta')+S_1(c||d).
}
Since $\Tr_C[\tau]$ and $\rho'$ are arbitrarily close, the above relation directly implies $S_1(\rho||\eta)\geq S_1(\rho'||\eta')$.

\bigskip

We now treat the difficult part: $1)\Rightarrow 2)$.
It suffices to prove the case without equality (i.e., $S_1(\rho||\eta)> S_1(\rho'||\eta')$) for the following reason:
In the case of $S_1(\rho||\eta)=S_1(\rho'||\eta')$, for any given $\ep>0$, there exists a state $\rho''$ such that $S_1(\rho||\eta)>S_1(\rho''||\eta')$ and $d_1(\rho', \rho'')<\ep/2$.
A simple construction of such a state is $\rho''=\( 1-\frac{\ep}{2}\) \rho'+\frac{\ep}{2} \eta'$.
In this choice, the condition $d_1(\rho', \rho'')<\ep/2$ is satisfied by construction, and the condition $S_1(\rho||\eta)=S_1(\rho'||\eta')>S_1(\rho''||\eta')$ is confirmed by the following fact:
The relative entropy is a strictly convex function with respect to the first argument (e.g., Exercise 11.18 of \cite{NCbook}).
Since $\rho''$ is an internal dividing point between $\rho'$ and $\eta'$, and $S(\rho'||\eta')>S(\eta'||\eta')=0$, we find that $S(\rho'||\eta')>S(\rho''||\eta')>0$.
By applying Theorem 3 without the equality with setting $\ep$ to $\ep/2$ and $\rho'$ to $\rho''$, we find that there exist a catalyst system $C$, its state $c$ and $d$, and a CPTP map $\calN$ on the composite system $SC$ such that $\calN(\rho\otimes c)=\tau$ and $\calN(\eta\otimes d)=\eta'\otimes d$, where $\tau$ satisfies $\Tr_S[\tau]=c$, $d_1(\Tr_C[\tau], \rho'')<\ep/2$, and $I_{SC}[\tau]<\delta$.
Combining  $d_1(\rho', \rho'')<\ep/2$ and $d_1(\Tr_C[\tau], \rho'')<\ep/2$, we conclude that $d_1(\Tr_C[\tau], \rho')<\ep$ and hence this CPTP map $\calN$ is the desired map.

We shall prove $1)\Rightarrow 2)$ for the case without equality in the following three steps.

\bigskip

\underline{Step 1: Deriving a sufficient condition for state conversion.}

\midskip

In the first step, we show a useful lemma.
This lemma employs two kinds of divergences: the quantum hypothesis testing divergence and the quantum R\'{e}nyi-$\infty$ divergence.

The quantum hypothesis testing divergence of the first kind is defined as
\eq{
\SH^{1-\ep} (\sigma||\kappa) :=-\ln \( \min_{0\leq Q\leq I, \ \Tr[\sigma Q]\geq 1-\ep} \Tr[\kappa Q] \) ,
}
where $I$ is the identity operator.
In quantum hypothesis testing, an unknown quantum state which takes $\sigma$ or $\kappa$ is provided, and the task is to identify it.
In our setup, the error of the first (resp. second) kind is the probability that when the actual state is $\sigma$(resp. $\kappa$), we incorrectly guess that the state is $\kappa$(reps. $\sigma$).
The quantum hypothesis testing divergence is the logarithm of the minimum error of the second kind under a fixed amount of the error of the first kind.
The operator $Q$ serves as a measurement operator for guessing that the state is $\sigma$ (and $1-Q$ for $\kappa$), and the condition $\Tr[\sigma Q]\geq 1-\ep$ implies that the error of the first kind is less than $\ep$ (i.e., if the state is $\sigma$, we guess that the state is $\sigma$ with probability at least $1-\ep$).
Under this condition, we minimize $\Tr[\kappa Q]$, which is the probability that we guess that the state is $\sigma$ when the actual state is $\kappa$.

We also introduce the quantum R\'{e}nyi-$\infty$ divergence $S_\infty (\sigma||\kappa)$ defined as
\eq{
S_\infty (\sigma||\kappa):=\ln (\min[\lmd: \sigma\leq \lmd \kappa]).
}
We sometimes use this quantity in the form $\sigma\leq e^{S_\infty (\sigma||\kappa)}\kappa$.
The following lemma plays a key role, which is Lemma 3.3 of Ref.~\cite{BST} (see also Refs.~\cite{FR, Fai, Sag, Sag-review, WW19}, which state almost the same result).

\bigskip

{\it Lemma 1}:
Suppose that the following relation
\eq{
S_{\rm H}^{1-\ep} (\sigma||\kappa) \geq S_\infty (\sigma'||\kappa') 
}
holds for a fixed $0<\ep<1$.
%Here, we set $q:=1-e^{-s'-\beta w}/Z$ with $Z:=1+e^{-\beta w}$.
%We denote $s:=S_{\rm H}^{\eta=1} (\sigma||\sgmG)$ and $s':=S_\infty (\sigma'||\sgmG)$ for brevity.
Then, there exists a CPTP map $\Lambda$ which satisfies
\balign{
\Lambda (\kappa)&=\kappa' , \\
d_1(\Lambda (\sigma), \sigma')&<\ep.
}

\bigskip

While the proof of this lemma is presented in Ref.~\cite{BST}, we reproduce it in order to make this Supplemental Material self-contained.

\begin{proof}[Proof of Lemma 1]
The proof of Lemma 1 is based on the measurement-and-prepare method.
In this proof, we employ the abbreviation $s:=S_{\rm H}^{1-\ep} (\sigma||\kappa)$ for brevity.

We first perform a quantum measurement.
By definition of the hypothesis testing divergence, there exists a nonnegative Hermitian operator $A$ satisfying $\Tr[A\sigma]=1-\ep$ and $\Tr[A\kappa]=e^{-s}$.
Thus, by performing the POVM measurement with $(A, 1-A)$ on $\sigma$ and $\kappa$, we map quantum states into classical probability distributions of two measurement outcomes:
\eq{
\sigma \to \mx{1-\ep \\ \ep}, \hspace{10pt} 
\kappa \to \mx{e^{-s} \\ 1-e^{-s}}.
}

We next prepare quantum states depending on the outcome of the measurement.
If the measurement outcome corresponds to $A$, we prepare $\sigma'$, and if the measurement outcome corresponds to $1-A$, we prepare 
\eq{
\kappa'':=\frac{\kappa'-e^{-s}\sigma'}{1-e^{-s}}.
}
The relation $s\geq S_\infty(\sigma'||\kappa')$ guarantees that $\kappa''$ is nonnegative.
Through this preparation, $\mx{1-\ep \\ \ep}$ is converted into $\sigma'':=(1-\ep)\sigma'+\ep \kappa''$, which satisfies $d_1(\sigma'', \sigma')\leq \ep$.
In a similar manner, $\mx{e^{-s}  \\ 1-e^{-s} }$ is converted into $\kappa'$.
We thus obtain the desired CPTP map.
\end{proof}

\underline{Step 2: Applying the quantum Stein's lemma.}

\midskip

We next apply the quantum Stein's lemma, which is one of the most important results in the field of quantum hypothesis testing.
Ogawa and Nagaoka~\cite{ON} proved that for multi-copy systems $\rho^{\otimes n}$ and $\eta^{\otimes n}$ any quantum hypothesis testing divergence rate with $0<\ep<1$ converges to the KL divergence in the infinite copy limit:
\eqa{
\lim_{n\to \infty}\frac 1n S_{\rm H}^{1-\ep} (\rho^{\otimes n}||\eta^{\otimes n})=S_1(\rho||\eta).
}{SH-KL}
This relation plays a crucial role to derive a single monotone in quantum thermodynamics (see also Ref.~\cite{Sag}).

Datta~\cite{Dat} found the connection between the quantum hypothesis testing divergence and the $\ep$-smoothed R\'{e}nyi divergence.
Define the $\ep$-neighborhood of $\rho$ as $B^\ep (\rho):=\{ \tilde{\rho}: d_1(\rho, \tilde{\rho})\leq \ep\}$, where $\tilde{\rho}$ is a density operator.
We define the $\ep$-smoothed R\'{e}nyi-$\infty$ divergence as
\eq{
S_\infty^\ep (\rho||\eta):=\min_{\tilde{\rho} \in B^\ep(\rho)} S_\infty (\tilde{\rho}||\eta).
}
By using the quantum Stein's lemma, Datta~\cite{Dat} showed that for any $0<\ep<\frac12$ the $\ep$-smoothed R\'{e}nyi-$\infty$ divergence rate also converges to the KL divergence in the infinite copy limit:
\eqa{
\lim_{n\to \infty}\frac 1n S_\infty^\ep  (\rho^{\otimes n}||\eta^{\otimes n})=S_1(\rho||\eta).
}{inf-KL}

Combining Eqs.\eqref{SH-KL}, \eqref{inf-KL}, and Lemma 1 with setting $\ep$ as $\ep/2$, we arrive at the following key result:
If $S_1(\rho||\eta)> S_1(\rho'||\eta')$ is satisfied, then for any $\ep>0$ there exists a sufficiently large $n$ such that there is a CPTP map $\Lambda$ which converts
\eq{
\Lambda(\eta^{\otimes n})={\eta'}^{\otimes n}
}
and
\eq{
\Lambda(\rho^{\otimes n})=\Xi
}
with $d_1(\rho'^{\otimes n}, \Xi)<\ep$.

\bigskip

\underline{Step 3: Reduction of asymptotic state conversion to catalytic conversion.}

\midskip

Our remaining task is to reduce the approximate asymptotic (multi-copy) state conversion to the catalytic state conversion.
To this end, we consider a refinement of the existing technique~\cite{Dua, AN, Bra15}.
In Refs.~\cite{Dua, AN}, they consider {\it exact} asymptotic state conversion and reduce it to catalytic conversion, while we consider {\it approximate} asymptotic state conversion and need to reduce it to catalytic conversion.

We set the Hilbert space of the catalyst system as $S^{\otimes n-1}\otimes A$, where $A$ is spanned by $\{ \ket{1}, \ket{2}, \ldots , \ket{n}\}$, and $n$ is the large integer introduced in Step 2.
We call the system $S$ also as $S_1$, and $n-1$ copies of the same systems in the catalyst as $S_2, \ldots, S_n$.
The remaining subsystem of the catalyst $A$ serves as a label.

Using $\Xi$ on $S_1\otimes S_2\otimes\cdots \otimes S_n$, we introduce $\Xi_i$ ($i\in \{ 1,\ldots ,n\}$) as a reduced state of $\Xi$ on $S_1\otimes \cdots \otimes S_i$.
Note that we later consider the shift of $\Xi_i$ in $S_1\otimes \cdots \otimes S_n$.
We interpret $\Xi_0$ as a trivial state $1$ for convenience.
Using these states, we set the state of the catalyst $c$ as 
\eq{
c:=\frac1n \sum_{k=1}^n \rho^{\otimes k-1}\otimes \Xi_{n-k}\otimes \ket{k}\bra{k},
}
where $\rho^{\otimes k-1}$ is on $S_2\otimes \cdots \otimes S_k$, and $\Xi_{n-k}$ is now on $S_{k+1}\otimes \cdots \otimes S_n$.
In a similar manner, we set $d$ as
\eq{
d:=\frac1n \sum_{k=1}^n \eta^{\otimes k-1}\otimes \eta'^{\otimes n-k}\otimes \ket{k}\bra{k}.
}

We now construct the desired CPTP map, which consists of the following three processes:
\be{
\renewcommand{\theenumi}{\Roman{enumi}}
\item If the auxiliary system $A$ is $\ket{n}\bra{n}$, then we apply $\Lambda$ (obtained in Step 2) to $S_1\otimes \cdots \otimes S_n$. 
Otherwise, we leave $S_1\otimes \cdots \otimes S_n$ as it is.
\item We convert the auxiliary system $A$ as $\ket{n}\to \ket{1}$ and $\ket{i}\to \ket{i+1}$.
\item We shift the systems as $S_i\to S_{i+1}$ and $S_n\to S_1$.
}
Through the processes I and II, the initial state
\eq{
\rho\otimes c=\frac1n \sum_{k=1}^n \rho^{\otimes k}\otimes \Xi_{n-k}\otimes \ket{k}\bra{k}
}
is converted into 
\eq{
\tau'=\frac1n \sum_{k=1}^n \rho^{\otimes k-1}\otimes \Xi_{n+1-k}\otimes \ket{k}\bra{k}.
}
We note that the partial trace of $\tau'$ with respect to $S_n$ recovers the initial state of the catalyst $c$.
We denote by $\tau$ the state after the shift process III on $\tau'$.
In a similar manner, through the processes I and II, the initial state
\eq{
\eta\otimes d=\frac1n \sum_{k=1}^n \eta^{\otimes k}\otimes \eta'^{\otimes n-k}\otimes \ket{k}\bra{k}
}
is converted into
\eqa{
\frac1n \sum_{k=1}^n \eta^{\otimes k-1}\otimes \eta'^{\otimes n-k+1}\otimes \ket{k}\bra{k}.
}{eta-d-last}
Through the shift process III, the above term \eqref{eta-d-last} becomes $\eta'\otimes d$.

We finally show that $\Tr_C[\tau]$ is $\ep$-close to $\rho'$ and the correlation is small.
We first treat the former.
The state $\Tr_C[\tau]$ is expressed as 
\eq{
\Tr_C[\tau]=\frac 1n \sum_{k=1}^n \xi_l,
}
where we defined $\xi_l$ as the reduced state of $\Xi$ on $S_l$: $\xi_l:=\Tr_{[1,2,\ldots , l-1,l+1,\ldots ,n]}[\Xi]$.
Then, using the monotonicity and convexity of the trace distance $d_1$, we arrive at the desired result
\eq{
d_1\( \frac 1n \sum_{l=1}^n \xi_l, \rho'\) \leq \frac1n \sum_{l=1}^n d_1(\xi_l, \rho') \leq  \frac1n \sum_{l=1}^n d_1(\Xi, \rho'^{\otimes n})<\ep .
}

We next prove the latter condition (i.e., $I_{SC}[\tau]<\delta$).
By construction, $\Xi$ can be written in the form of $\Xi=(1-\ep)\rho'^{\otimes n}+\ep X$, and hence $\tau$ is written as
\eq{
\tau=(1-\ep) \rho'\otimes \( \frac1n \sum_{k=1}^n \rho^{\otimes k}\otimes \rho'^{\otimes n-k-1}\otimes \ket{k}\bra{k} \) +\sum_{i=1}^d \ep_i \rho_i\otimes c_i,
}
where $\sum_{i=1}^d \ep_i=\ep$, $d$ is the dimension of the Hilbert space of $S$, and $\rho_i$ and $c_i$ are some states of $S$ and $C$, respectively.
Then, using the subadditivity and the Araki-Lieb inequality for the von Neumann entropy~\cite{NCbook}, the correlation between the system and the catalyst is evaluated as
\eq{
I_{SC}[\tau]\leq 2\[ -(1-\ep)\ln (1-\ep) -\sum_{i=1}^d \ep_i \ln \ep_i\] \leq 2\[ -(1-\ep)\ln (1-\ep) - \ep \ln \frac{\ep}{d} \] .
}
Thus, by replacing $\ep$ to $\tilde{\ep}$ such that $2\[ -(1-\tilde{\ep})\ln (1-\tilde{\ep}) - \tilde{\ep} \ln \tilde{\ep}/{d} \] \leq \delta$, we obtain the desired inequality $I_{SC}[\tau]<\delta$.
This completes the proof of Theorem 3.
\end{proof}

%%%%%%%%%%%%%%%%%%%%%%%%%%%%%%%%%%%%%%%
\bigskip\noindent
{\bf \large B. Proof of Theorem 2}
\midskip
%%%%%%%%%%%%%%%%%%%%%%%%%%%%%%%%%%%%%%%

We first present the rigorous statement of Theorem 2:

\bigskip

{\it Theorem 2}--
Consider two states $\rho$ and $\rho'$ of the system $S$ with $F(\rho)-F(\rho')<0$.
Then, the following two conditions are equivalent.
\be{
\renewcommand{\theenumi}{\arabic{enumi})}
\item $F(\rho)-F(\rho')\geq \beta w$.
\item For any $t>0$, $u>0$, and $\delta>0$, there exist a catalyst $C$, its state $c$, a work storage $W$ with two eigenstates $\ket{a}$, $\ket{b}$ with energies $E_a=0$ and $E_b=w<0$, and a Gibbs-preserving map $\rho \otimes c \otimes \ket{a}\bra{a} \to \tau \otimes \omega$ such that (i) $\Tr_S[\tau]=c$, (ii) $d_1(\Tr_C[\tau], \rho')<t$, (iii) $d_1(\omega, \ket{b}\bra{b})<u$, and (iv) $I_{SC}[\tau]<\delta$.
}

\bigskip

\begin{proof}

The proof of $2)\Rightarrow 1)$ is straightforward as in the case of Theorem 3.
The additivity and the monotonicity of the KL divergence imply that
\balign{
S_1(\rho||\rhoG)+S_1(c||\cG)+S_1(\ket{a}\bra{a}||\omgG)=&S_1(\rho\otimes c\otimes \ket{a}\bra{a}||\rhoG\otimes \cG\otimes \omgG ) \nt \\
\geq& S_1(\tau\otimes \omega||\rhoG\otimes \cG\otimes \omgG) \nt \\
\geq& S_1(\Tr_C[\tau]||\rhoG)+S_1(c||\cG)+S_1(\omega||\omgG),
}
where $\cG$ is the Gibbs state of $C$, and we defined
\eq{
\omgG:=\frac{1}{z}\ket{a}\bra{a}+\frac{e^{-\beta w}}{z}\ket{b}\bra{b}
}
with $z:=1+e^{-\beta w}$ is the Gibbs state of $W$.
Since $\omega$ is arbitrarily close to $\ket{b}\bra{b}$ and $S_1(\ket{b}\bra{b}||\omgG)-S_1(\ket{a}\bra{a}||\omgG)=\beta w$, we arrive at the desired relation $F(\rho)-F(\rho')\geq \beta w$.

\bigskip

We now prove $1)\Rightarrow 2)$.
In order to apply the reduction from asymptotic state conversion to catalytic conversion (Step 3 of Theorem 3), we need the following lemma, which serves as a combination of Step 1 and Step 2:
% to modify the result in the previous section (Lemma 2) to the case of multiple work storages.
%It is not easy but we fortunately succeed in constructing a protocol in the case of the work cost (i.e., $w<0$).
%This result is not applicable to the work extraction (i.e., $w>0$) at present.

%The crucial fact in the proof of Theorem 2 is the following lemma, which serves as the combination of the Step 1 and Step 2 in the proof of Theorem 3:

\bigskip

{\it Lemma 2}:
Suppose that the following relation holds
\eq{
S_1(\rho||\rhoG)>S_1 (\rho'||\rhoG)+\beta w.
}
Then, for any $t>0$ and $u>0$, there exists an integer $m$ and a Gibbs-preserving map such that
\eq{
\rho^{\otimes m}\otimes \ket{a}\bra{a} ^{\otimes m} \to \Xi \otimes \omega ^{\otimes m}
}
with 
\balign{
d_1(\Xi, \rho'^{\otimes m})<&t,\\
d_1(\omega, \ket{b}\bra{b})<&u,
}
where we set $E_a=0$ and $E_b=w<0$.

\bigskip

We note that, for the same reason as the proof of Theorem 3, we only have to consider the case of $S_1(\rho||\rhoG)>S_1 (\rho'||\rhoG)+\beta w$ (without the equality case).

%Owing to the quantum Stein's lemma, for any $1/2>\ep>0$ there exists a sufficiently large $M$ such that for any $n>M$ the following relation
%\eq{S_{\rm H}^{1-\ep} (\rho^{\otimes n}||\rhoG^{\otimes n}) \geq S_\infty (\rho'^{\otimes n}||\rhoG^{\otimes n}) + \beta n w.}

\begin{proof}[Proof of Lemma 2]
Without loss of generality, we suppose that $t<\frac12$ and $u<\min( e^{2w}, \frac{e^w}{2})$.
We set $\ep=\frac t2$ and $m=\max(\frac{2e^{2w}}{u}, M, 2)$, where $M$ is a sufficiently large integer such that $m\geq M$ implies
\eq{
S_{\rm H}^{1-\ep} (\rho^{\otimes m}||\rhoG^{\otimes m}) \geq S_\infty (\rho'^{\otimes m}||\rhoG^{\otimes m}) + \beta m w .
}
The quantum Stein's lemma guarantees the existence of such an integer $M$.
We denote $\sigma:=\rho^{\otimes m}$, $\sigma':=\rho'^{\otimes m}$, $\sgmG:=\rhoG^{\otimes m}$, and $s:=S_{\rm H}^{1-\ep} (\sigma||\sgmG)$, $s':=S_\infty (\sigma'||\sgmG)$ for brevity.
%The proof is very similar to that of Lemma 2.

We first perform not binary but ternary measurement on the composite system.
By definition of the hypothesis testing divergence, there exists a Hermitian operator $A$ satisfying $\Tr[A\sigma]=1-\ep$ and $\Tr[A\sgmG]=e^{-s}$.
Using this operator, we construct the POVM measurement with $\{ A\otimes \ket{a}\bra{a}^{\otimes m}, (1-A)\otimes \ket{a}\bra{a}^{\otimes m}, 1\otimes (1- \ket{a}\bra{a}^{\otimes m})\}$, for which we denote the measurement output by 1, 2, and 3, respectively.
Then, if the state is $\sigma\otimes \ket{a}\bra{a}^{\otimes m}$, the measurement outcome is 1 with probability $1-\ep$, 2 with probability $\ep$, and 3 with probability 0.
If the state is $\sgmG \otimes \omgG^{\otimes m}$, the measurement outcome is 1 with probability ${e^{-s}}\frac{1}{z^m}$, 2 with probability $(1-e^{-s})\frac{1}{z^m}$, and 3 with probability $1-\frac{1}{z^m}$.

%Here, $x$ runs all $2^m$ sequences of $\{a,b\}$ with length $m$, from $\ket{aa\cdots a}$ to $\ket{bb\cdots b}$, and $N_a(x)$ ($N_b(x)$) is the number of $a$($b$) in the sequence $x$.

Then, we prepare quantum states based on the measurement outcome.
If the measurement output is $i$ ($i= 1, 2, 3$), we prepare a quantum state as
\eq{
\zeta^{i}:=\sum_{y\in \{ a,b\}^{\otimes m}} \[ v^{i}_{1y} \sigma' \otimes \ket{y}\bra{y}+v^{i}_{0y} \sgmG \otimes \ket{y}\bra{y} \],
}
where $y$ runs $2^m$ possible sequences of $\{ a,b\}^{\otimes m}$ and $v^i_{jy}$'s are coefficients.
The normalization condition requires $\sum_{jy}v^{i}_{jy}=1$ for any $i$.
In addition, $\zeta^i$ is positive-semidefinite if
\eqa{
v^{i}_{0y}e^{-s'}+v^{i}_{1y}\geq 0
}{cond-sinf}
holds for any $i$ and $y$.

We set $v^i_{jy}$'s as 
\balign{
v^{1}_{1y}&=\frac{1-t}{1-\ep}u^{N_a(y)}(1-u)^{N_b(y)}, \\
v^{1}_{0y}&=\frac{t-\ep}{1-\ep}u^{N_a(y)}(1-u)^{N_b(y)}, \\
v^{2}_{1y}&=0, \\
v^{2}_{0y}&=u^{N_a(y)}(1-u)^{N_b(y)}, \\
v^3_{1y}&=-\frac{1}{z^m-1}e^{-s}\frac{1-t}{1-\ep}u^{N_a(y)}(1-u)^{N_b(y)}, \\
v^3_{0y}&=\frac{1}{z^m-1}\[ e^{-wN_b(y)}-e^{-s}\frac{t-\ep}{1-\ep}u^{N_a(y)}(1-u)^{N_b(y)}-(1-e^{-s})u^{N_a(y)}(1-u)^{N_b(y)}\] , 
}
where $N_a(y)$ (resp. $N_b(x)$) is the number of $a$ (resp. $b$) in the sequence $y$.
Remarkably, these $v^i_{jy}$'s satisfy
\balign{
(1-\ep)v^{1}_{1y}+\ep v^{2}_{1y}=&(1-t)u^{N_a(y)}(1-u)^{N_b(y)}, \lb{cond1-1} \\
(1-\ep)v^{1}_{0y}+\ep v^{2}_{0y}=&tu^{N_a(y)}(1-u)^{N_b(y)}, \lb{cond1-2}
}
and 
\balign{
\frac{e^{-s}}{z^m}v^1_{1y}+ \frac{1-e^{-s}}{z^m}v^2_{1y}+\( 1-\frac{1}{z^m}\) v^3_{1y}&=0, \lb{cond2-1} \\
\frac{e^{-s}}{z^m}v^1_{0y}+ \frac{1-e^{-s}}{z^m}v^2_{0y}+\( 1-\frac{1}{z^m}\) v^3_{0y}&=\frac{e^{-wN_b(y)}}{z^m}, \lb{cond2-2}
}
for any $y$, which implies that this measurement-and-preparation CPTP map converts
\balign{
\sigma\otimes \ket{a}\bra{a}^{\otimes m} &\to ((1-t)\sigma'+t \sgmG)\otimes (u\ket{a}\bra{a}+(1-u)\ket{b}\bra{b})^{\otimes m}, \\
\sgmG\otimes \omgG^{\otimes m} &\to \sgmG\otimes \omgG^{\otimes m}. \lb{Gibbs-return-Thm2}
}
Namely, the constructed CPTP map is the desired Gibbs-preserving map.
Note that we chose the values of $v^i_{jy}$'s in the following procedure:
We first set $v^{2}_{1y}=0$, and then require the strengthened normalization condition $v^i_{1y}+v^i_{0y}=u^{N_a(y)}(1-u)^{N_b(y)}$ ($i=1,2,3$) and Eqs.~\eqref{cond1-1}, \eqref{cond1-2}, \eqref{cond2-1}, \eqref{cond2-2}, which uniquely determine the other five values.

We finally demonstrate that these $v^i_{jy}$'s satisfy the positive-semidefinite condition (\eref{cond-sinf}).
Since the cases of $i=1,2$ are trivial, below we only treat the case of $i=3$.
The positive-semidefinite condition $v^{3}_{0y}e^{-s'}+v^{3}_{1y}\geq0$ can be transformed as
\balign{
{e^{-wN_b(y)}}\geq e^{s'-s}\frac{1-t}{1-\ep}u^{N_a(y)}(1-u)^{N_b(y)}+e^{-s}\frac{t-\ep}{1-\ep}u^{N_a(y)}(1-u)^{N_b(y)}
&+(1-e^{-s})u^{N_a(y)}(1-u)^{N_b(y)}.
}
Owing to $s\geq s'+mw$ and $N_a(y)+N_b(y)=m$, it suffices to prove
\balign{
1\geq &e^{-N_a(y)w}\frac{1-t}{1-\ep}u^{N_a(y)}(1-u)^{N_b(y)}+e^{-N_a(y)w-s'}\frac{t-\ep}{1-\ep}u^{N_a(y)}(1-u)^{N_b(y)} +e^{N_b(y)w}(1-e^{-s})u^{N_a(y)}(1-u)^{N_b(y)} \lb{cond-sinf-fin}
}
for any $y$.
%Note that $w<0$.
Finally, using $e^{2w}/m<u<e^{2w}$, $u<e^w/2\leq 1/(1+e^{-w})$, and $s'\geq0$, we prove \eref{cond-sinf-fin} as
\balign{
&e^{-N_a(y)w}\frac{1-t}{1-\ep}u^{N_a(y)}(1-u)^{N_b(y)}+e^{-N_a(y)w-s'}\frac{t-\ep}{1-\ep}u^{N_a(y)}(1-u)^{N_b(y)}+e^{N_b(y)w}(1-e^{-s})u^{N_a(y)}(1-u)^{N_b(y)} \nt \\
\leq & (1+e^{mw})e^{-N_a(y)w}u^{N_a(y)}(1-u)^{N_b(y)} \nt \\
\leq & (1+e^{mw})(1-u)^m \nt \\
\leq & (1+e^{mw})\( 1-\frac{e^{2w}}{m}\) ^m \nt \\
< & (1+e^{2w})\frac{1}{e^{(e^{2w})}} \nt \\
\leq & (1+e^{2w})\frac{1}{1+e^{2w}} \nt \\
=&1.
}
In the second line we used $e^{-s'}\leq 1$ and $N_a(y)+N_b(y)=m$, in third line we used $e^{-w}u<\frac12 <1-u$, and in the fifth line we used $m\geq 2$ and $(1-a/m)^m<1/e^a$.
This completes the proof.
\end{proof}

Lemma 2 serves as the counterpart of Step 1 and Step 2 of the proof of Theorem 3.
By combining the idea of Step 3 of Theorem 3, we prove Theorem 2 as follows.

We set the catalyst as
\eq{
c:=\frac1m \sum_{k=1}^m \rho^{\otimes k-1}\otimes \Xi_{m-k}\otimes \ket{a}\bra{a}^{\otimes k-1}\otimes \omega^{\otimes m-k}\otimes \ket{k}\bra{k},
}
where we set $\Xi=(1-t)\rho'^{\otimes m}+t \rhoG^{\otimes m}$ and $\omega=u\ket{a}\bra{a}+(1-u)\ket{b}\bra{b}$, and the definition of $\Xi_{m-k}$ is the same as the proof of Theorem 3.
The choices of $t$, $u$, and $m$ are the same as Lemma 2.

We follow Step 3 of the proof of Theorem 3 by setting the Gibbs preserving map to that derived in Lemma 2 and replacing $S$ with $S\otimes W$.
Then, the initial state of the composite system $SWC$ is
\eq{
\rho\otimes c\otimes \ket{a}\bra{a}=\frac1m \sum_{k=1}^m \rho^{\otimes k}\otimes \Xi_{m-k}\otimes \ket{a}\bra{a}^{\otimes k}\otimes \omega^{\otimes m-k}\otimes \ket{k}\bra{k}
}
and the final state is
\eq{
\frac1m \sum_{k=1}^m \rho^{\otimes k-1}\otimes \Xi_{m-k+1}\otimes \ket{a}\bra{a}^{\otimes k-1}\otimes \omega^{\otimes m-k+1}\otimes \ket{k}\bra{k}.
}
We thus have
\eq{
\calN(\rho\otimes c\otimes \ket{a}\bra{a})=\tau \otimes \omega 
}
with $\Tr_S[\tau]=c$, $d_1(\Tr_C[\tau], \rho')<t$, $d_1(\omega, \ket{b}\bra{b})<u$.
In addition, by setting $t$ sufficiently small if needed, we obtain $I_{SC}[\tau]<\delta$.
\end{proof}

%%%%%%%%%%%%%%%%%%%%%%%%%%%%%%%%%%%%%%%
\bigskip\noindent
{\bf \large C. Sufficient condition for work investment/extraction without catalyst}
\midskip
%%%%%%%%%%%%%%%%%%%%%%%%%%%%%%%%%%%%%%%

As a side remark, we show an interesting lemma: a necessary condition for state conversion with a two-level work storage $W$ with states $a$ and $b$.
Unfortunately, this lemma does not directly apply to our setup of Theorem 2.
However, this proof method inspires the proof of Theorem 2, and this lemma itself has potential applications to quantum thermodynamics.
In particular, this lemma applies in the case of work extraction.

\bigskip

{\it Lemma 3}:
Let $\sgmG$ be the Gibbs state of the system $S$ and consider two states $\sigma$ and $\sigma'$ of $S$.
Suppose that
\eq{
S_{\rm H}^{1-\ep} (\sigma||\sgmG) \geq S_\infty (\sigma'||\sgmG) + \beta w ,
}
where $\ep>0$ is sufficiently small (a more detailed condition is given soon later).
We denote $s:=S_{\rm H}^{1-\ep} (\sigma||\sgmG)$ and $s':=S_\infty (\sigma'||\sgmG)$ for brevity, and define $Z:=1+e^{-\beta w}$ and $q:=1-e^{-s'-\beta w}/Z$.
We assume that $\ep$ satisfies $0<\ep<\min (\frac12, \frac{4q^5}{(1+q)^2})$.

We claim that there exist a two-level system called work storage $W$ with two eigenstates $\ket{a},\ \ket{b}$ of the Hamiltonian with energies $E_a=0$ and $E_b=w$ and a Gibbs-preserving map on the composite system $SW$ such that it converts 
\eq{
\sigma\otimes \ket{a}\bra{a}\to \tlsgm\otimes \Omega
}
with 
\balign{
d_1(\tlsgm, \sigma')<&2\sqrt{\frac{\ep}{q}}, \\
d_1(\Omega, \ket{b}\bra{b})<&2\sqrt{\frac{\ep}{q}}.
}

\bigskip

The proof is very similar to that of Lemma 1 and Lemma 2.
We construct the desired Gibbs-preserving map by employing the measurement-and-preparation method as follows.

\begin{proof}
We write the Gibbs state of the work storage $W$ as 
\eq{
\OmgG:=\frac{1}{Z}\ket{a}\bra{a}+\frac{e^{-\beta w}}{Z}\ket{b}\bra{b}.
}

By definition of the hypothesis testing divergence, there is a nonnegative Hermitian operator $A$ satisfying $\Tr[A\sigma]=1-\ep$ and $\Tr[A\sgmG]=e^{-s}$.
Using this operator, we first perform a POVM measurement with $\{ A\otimes \ket{a}\bra{a}, 1-A\otimes \ket{a}\bra{a}\}$ on the composite system of the system and the work storage, which maps
\eq{
\sigma\otimes \ket{a}\bra{a} \to \mx{1-\ep \\ \ep}, \hspace{10pt} 
\sgmG \otimes \OmgG \to \mx{\frac{e^{-s}}{Z} \\ 1-\frac{e^{-s}}{Z}}.
}

We then apply a classical stochastic map $M$ such that
\eq{
\mx{1-\ep' \\ \ep'}=M\mx{1-\ep \\ \ep}, \hspace{10pt} 
\mx{\frac{e^{-s'-\beta w}}{Z}  \\ 1-\frac{e^{-s'-\beta w}}{Z} }=M\mx{\frac{e^{-s}}{Z} \\ 1-\frac{e^{-s}}{Z}}
}
with $\ep'\leq \ep$, which is realized by
\eq{
M:=\mx{1& 1- \frac{Z- e^{-s'-\beta w}}{Z-e^{-s}} \\
0& \frac{Z- e^{-s'-\beta w}}{Z-e^{-s}} 
} .
}
The condition $s\geq s'+\beta w$ guarantees the nonnegativity of matrix elements.

We now consider the preparation step.
By defining $\pwG:=e^{-\beta w}/Z$ and $u:=\sqrt{\ep/q}$, we introduce two states $\zeta_1$ and $\zeta_2$ as
\balign{
\zeta_1:=&\frac{(1-u)uq}{q-\ep'}\sigma'\otimes \ket{a}\bra{a} +\frac{(1-u)^2q}{q-\ep'}\sigma'\otimes \ket{b}\bra{b} \nt \\
&+\frac{u^2q-(1-\pwG)\ep'}{q-\ep'}\sgmG \otimes \ket{a}\bra{a} +\frac{u(1-u)q-\pwG\ep'}{q-\ep'}\sgmG\otimes \ket{b}\bra{b}, \\
\zeta_2:=&-\frac{(1-q)(1-u)u}{q-\ep'}\sigma'\otimes \ket{a}\bra{a} -\frac{(1-q)(1-u)^2}{q-\ep'}\sigma'\otimes \ket{b}\bra{b} \nt \\
&+\frac{(1-\ep')(1-\pwG)-(1-q)u^2}{q-\ep'}\sgmG \otimes \ket{a}\bra{a} +\frac{(1-\ep')\pwG-(1-q)u(1-u)}{q-\ep'}\sgmG\otimes \ket{b}\bra{b}.
}
Using these states, we prepare a quantum state from a classical distribution as
\eq{
\mx{p_1\\ p_2} \to p_1 \zeta_1+p_2 \zeta_2 .
}
This preparation realizes the desired state conversion:
\balign{
\mx{1-\ep' \\ \ep'}& \to ((1-u)\sigma'+u\sgmG )\otimes ((1-u)\ket{b}\bra{b}+u\ket{a}\bra{a}), \\
\mx{\frac{e^{-s'-\beta w}}{Z}  \\ 1-\frac{e^{-s'-\beta w}}{Z} }=\mx{1-q\\ q} &\to \sgmG \otimes \OmgG .
}

We now examine the conditions $0\leq \zeta_1\leq 1$ and $0\leq \zeta_2 \leq 1$.
We first consider $\zeta_1$.
It suffices to confirm the following inequalities:
\balign{
\frac{u^2q-(1-\pwG)\ep'}{q-\ep'}&\geq 0, \lb{zeta1-1} \\
\frac{u(1-u)q-\pwG\ep'}{q-\ep'}&\geq 0. \lb{zeta1-2}
}
Since $\frac{u(1-u)q-\pwG\ep'}{q-\ep'}\geq \frac{u^2q-(1-\pwG)\ep'}{q-\ep'}$, we only have to show \eref{zeta1-1}, which is easily verified as
\eq{
u^2q-(1-\pwG)\ep'\geq u^2q-\ep'\geq u^2q-\ep=0.
}
We next treat $\zeta_2$.
It suffices to confirm the following inequalities:
\balign{
\frac{(1-\ep')(1-\pwG)-(1-q)u^2}{q-\ep'}\sgmG-\frac{(1-q)(1-u)u}{q-u}\sigma'\geq &0, \\
\frac{(1-\ep')\pwG-(1-q)u(1-u)}{q-\ep'}\sgmG -\frac{(1-q)(1-u)^2}{q-u}\sigma' \geq & 0. \lb{zeta2-result}
}
We only have to show the latter one (\eref{zeta2-result}) because
\eq{
\frac{(1-\ep')(1-\pwG)-(1-q)u^2}{q-\ep'}\sgmG-\frac{(1-q)(1-u)u}{q-u}\sigma' \geq  \frac{(1-\ep')\pwG-(1-q)u(1-u)}{q-\ep'}\sgmG -\frac{(1-q)(1-u)^2}{q-u}\sigma' .
}
To verify \eref{zeta2-result}, we show the following inequality:
\balign{
\frac{q-\ep'}{q-u}\frac{(1-q)(1-u)^2}{(1-\ep')\pwG-(1-q)u(1-u)}=&e^{-s'}\frac{q-\ep'}{q-u}\frac{(1-q)(1-u)}{1-\ep'-e^{-s'}u(1-u)} \nt \\
\leq& e^{-s'}\frac{q-\ep'}{q-u}\frac{(1-q)(1-u)}{1-\ep'-u(1-u)} \nt \\
\leq& e^{-s'}\frac{q}{q-u}\frac{1-q}{1-u} \nt \\
\leq & e^{-s'}.
}
In the first line we used $\frac{1-q}{\pwG}=e^{-s'}$, in the second line we used $e^{-s'}\leq 1$, and in the third line we used $0<\ep'\leq \ep=u^2q\leq u$.
In the last line, we used $q(1-q)\leq (q-u)(1-u)$ for $u\leq \frac{2q^2}{1+q}<\frac{1+q-\sqrt{1+2q-3q^2}}{2}$ and $u^2=\frac{\ep}{q}$.
Using $\sgmG - e^{-s'} \sigma'\geq 0$, which comes from the definition of $s'=S_\infty (\sigma'||\sgmG)$, we arrive at the desired result:
\balign{
&\frac{(1-\ep')\pwG-(1-q)u(1-u)}{q-\ep'}\sgmG -\frac{(1-q)(1-u)^2}{q-u}\sigma' \nt \\
=&\frac{(1-\ep')\pwG-(1-q)u(1-u)}{q-\ep'}\( \sgmG -\frac{q-\ep'}{q-u}\frac{(1-q)(1-u)^2}{(1-\ep')\pwG-(1-q)u(1-u)}\sigma'\) \nt \\
\geq&\frac{(1-\ep')\pwG-(1-q)u(1-u)}{q-\ep'}(\sgmG - e^{-s'} \sigma') \nt \\
\geq &0.
}

\end{proof}

%%%%%%%%%%%%%%%%%%%%%%%%%%%%%%%%%%%%%%%
\bigskip\noindent
{\bf \large D. Details of the toy example in the main text}
\midskip
%%%%%%%%%%%%%%%%%%%%%%%%%%%%%%%%%%%%%%%

We here prove 
\eqa{
\rhoG^{\otimes 8}-\lmd \ket{+}\bra{+}^{\otimes 8}\geq 0
}{eq-IV}
for $\lmd\leq \( \frac{3}{8}\) ^8$.
We first decompose the energy eigenstates of the Hamiltonian $H=\( \frac34 \ket{0}\bra{0}+\frac14 \ket{1}\bra{1}\) ^{\otimes 8}$ as
\eq{
\ket{E_i}=a_i \ket{+}^{\otimes 8}+b_i \ket{X},
}
where $\ket{X}$ is some state of 8 two-level systems orthogonal to $\ket{+}^{\otimes 8}$.
Since all of $2^8$ computational basis states $\ket{00\cdots 00}$, $\ket{00\cdots 01}$, $\ldots$, $\ket{11\cdots 11}$ are energy eigenstates of $H$, we find that $a_i=\frac{1}{2^{8/2}}=\frac{1}{16}$ for any $i$.

Note that maximum of $\lmd$ satisfying \eref{eq-IV}, which we denote by $\lmd^*$, is expressed as
\eqa{
\lmd^*=\min_{\ket{\phi}}\frac{\braket{\phi|\rhoG^{\otimes 8}|\phi}}{\braket{\phi|\( \ket{+}\bra{+}^{\otimes 8}\) |\phi}}=\min_{c_i}\frac{\sum_i \abs{c_i}^2p_i^{\rm G}}{(\sum_i  \abs{c_ia_i})^2}=\frac{\sum_i \abs{c^*_i}^2p_i^{\rm G}}{(\sum_i  \abs{c^*_ia_i})^2},
}{lmd*-mid}
where $\ket{\phi}=\sum_i c_i \ket{E_i}$ runs all possible states spanned by $\{ \ket{0}, \ket{1}\} ^{\otimes 8}$, and $c^*_i$ is the optimal choice of this minimization.
In addition, $p_i^{\rm G}$ is the Boltzmann weight of the energy eigenstate $\ket{E_i}$ given by $p_i^{\rm G}=3^{N_i}/4^8$, where $N_i$ is the number of $\ket{0}$'s in the energy eigenstate $\ket{E_i}$.
Applying the Schwarz inequality to the right-hand side of \eref{lmd*-mid}, we have
\eq{
\lmd^*=\frac{\sum_i \abs{c^*_i}^2p_i^{\rm G}}{(\sum_i  \abs{c^*_ia_i})^2}\geq \frac{1}{\sum_i \frac{\abs{a_i}^2}{p_i^{\rm G}}}=\frac{1}{2^8 \sum_i 3^{-N_i}}=\( \frac38\) ^8.
}
This relation directly implies that $\rhoG^{\otimes 8}-\lmd \ket{+}\bra{+}^{\otimes 8}\geq 0$ for $\lmd\leq \( \frac38\) ^8$.

\end{widetext}

\end{document}